\documentclass[]{spie}  

\usepackage{amsmath,amsfonts,amssymb}
\usepackage{graphicx}
\usepackage[colorlinks=true, allcolors=blue]{hyperref}
\usepackage{multirow}
\usepackage{wasysym}
\usepackage[perpage]{footmisc}

\title{The Simons Observatory: the Large Aperture Telescope Receiver (LATR) Integration and Validation Results}

\author[1, 2]{Zhilei Xu}
\author[1]{Tanay Bhandarkar}
\author[1, 3]{Gabriele Coppi}
\author[1]{Anna M.~Kofman}
\author[1]{John L.~Orlowski-Scherer}
\author[1]{Ningfeng Zhu}

\author[4]{Aamir M.~Ali}
\author[5]{Kam Arnold}
\author[6]{Jason E.~Austermann}
\author[7, 8]{Steve K.~Choi}
\author[9, 6]{Jake Connors}
\author[10]{Nicholas F.~Cothard}
\author[1]{Mark Devlin}
\author[1]{Simon Dicker}
\author[9, 6]{Bradley Dober}
\author[6]{Shannon M.~Duff}
\author[11]{Giulio Fabbian}
\author[5]{Nicholas Galitzki}
\author[1]{Saianeesh K.~Haridas}
\author[12]{Kathleen Harrington}
\author[13]{Erin Healy}
\author[14]{Shuay-Pwu Patty Ho}
\author[6]{Johannes Hubmayr}
\author[1]{Jeffrey Iuliano}
\author[15]{Jack Lashner}
\author[7]{Yaqiong Li}
\author[1]{Michele Limon}
\author[16]{Brian J.~Koopman}
\author[13]{Heather McCarrick}
\author[17]{Jenna Moore}
\author[3]{Federico Nati}
\author[7, 8]{Michael D.~Niemack}
\author[18]{Christian L.~Reichardt}
\author[1]{Karen Perez Sarmiento}
\author[5]{Joseph Seibert}
\author[5]{Maximiliano Silva-Feaver}
\author[13]{Rita F.~Sonka}
\author[13]{Suzanne Staggs}
\author[19, 1]{Robert J.~Thornton}
\author[7]{Eve M.~Vavagiakis}
\author[6]{Michael R.~Vissers}
\author[9, 6]{Samantha Walker}
\author[13]{Yuhan Wang}
\author[20]{Edward J.~Wollack}
\author[13]{Kaiwen Zheng}

\affil[1]{Department of Physics and Astronomy, University of Pennsylvania, Philadelphia, PA, USA}
\affil[2]{MIT Kavli Institute, Massachusetts Institute of Technology, Cambridge, MA, USA}
\affil[3]{Department of Physics, University of Milano-Bicocca, Milano (MI), Italy}
\affil[4]{Department of Physics, University of California, Berkeley, CA, USA}
\affil[5]{Department of Physics, University of California San Diego, La Jolla, CA, USA}
\affil[6]{National Institute of Standards and Technology, Boulder, CO, USA}
\affil[7]{Department of Physics, Cornell University, Ithaca, NY, USA}
\affil[8]{Department of Astronomy, Cornell University, Ithaca, NY, USA}
\affil[9]{Department of Physics, University of Colorado Boulder, Boulder, CO, USA}
\affil[10]{Department of Applied and Engineering Physics, Cornell University, Ithaca, NY, USA}
\affil[11]{Department of Physics \& Astronomy, University of Sussex, Brighton, UK}
\affil[12]{Department of Astronomy and Astrophysics, University of Chicago, Chicago, IL, USA}
\affil[13]{Department of Physics, Princeton University, Princeton, NJ, USA}
\affil[14]{Department of Physics, Stanford University, CA, USA}
\affil[15]{Department of Physics and Astronomy, University of Southern California, LA, CA, USA}
\affil[16]{Department of Physics, Yale University, New Haven, CT, USA}
\affil[17]{School of Earth and Space Exploration, Arizona State University, Tempe, AZ, USA}
\affil[18]{School of Physics, University of Melbourne, Parkville, Australia}
\affil[19]{Department of Physics, West Chester University of Pennsylvania, West Chester, PA, USA}
\affil[20]{NASA/Goddard Space Flight Center, Greenbelt, MD, USA}

\authorinfo{Corresponding author: \href{mailto:zhileixu@sas.upenn.edu}{zhileixu@sas.upenn.edu}}

\begin{document} 
\maketitle

\begin{abstract}
The Simons Observatory (SO) will observe the cosmic microwave background (CMB) from Cerro Toco in the Atacama Desert of Chile. The observatory consists of three 0.5\,m Small Aperture Telescopes (SATs) and one 6\,m Large Aperture Telescope (LAT), covering six frequency bands centering around 30, 40, 90, 150, 230, and 280\,GHz. The SO observations will transform the understanding of our universe by characterizing the properties of the early universe, measuring the number of relativistic species and the mass of neutrinos, improving our understanding of galaxy evolution, and constraining the properties of cosmic reionization.\cite{sofc19} As a critical instrument, the Large Aperture Telescope Receiver (LATR) is designed to cool $\sim$\,60,000 transition-edge sensors (TES)\cite{irwin/hilton:2005} to $<$\,100\,mK on a 1.7\,m diameter focal plane. The unprecedented scale of the LATR drives a complex design\cite{zhu/etal:2018, orlowski/etal:2018, coppi/etal:2018}. In this paper, we will first provide an overview of the LATR design. Integration and validation of the LATR design are discussed in detail, including mechanical strength, optical alignment, and cryogenic performance of the five cryogenic stages (80\,K, 40\,K, 4\,K, 1\,K, and 100\,mK). We will also discuss the microwave-multiplexing ($\mu$Mux) readout system implemented in the LATR and demonstrate the operation of dark prototype TES bolometers. The $\mu$Mux readout technology enables one coaxial loop to read out $\mathcal{O}(10^3)$ TES detectors. Its implementation within the LATR serves as a critical validation for the complex RF chain design. The successful validation of the LATR performance is not only a critical milestone within the Simons Observatory, it also provides a valuable reference for other experiments, e.g. CCAT-prime\cite{vavagiakis/etal:2018} and CMB-S4\cite{s4tb17, s4sb16}. 
\end{abstract}


\section{INTRODUCTION}
\label{sec:intro} 
In the past decades, the cosmic microwave background (CMB) observations have established the $\Lambda$CDM cosmology model;\cite{planck:2020, bennett/etal:2013, fixsen/etal:1996} recent advancement of ground-based CMB observations\cite{thornton/etal:2016, Ruhl2004} have continued to constrain the cosmological parameters and become powerful probes for astrophysical studies as well. The Simons Observatory (SO)\cite{galitzki/etal:2018} is a new ground-based CMB experiment in the Atacama Desert in Chile, comprised of three Small Aperture Telescopes (SATs)\cite{ali/etal:2020, kiuchi/etal:2020} and one Large Aperture Telescope (LAT). SO observes the CMB polarization at six frequency bands centering around 30, 40, 90, 150, 230, and 280\,GHz. The SATs will target the largest angular scales, mapping $\sim$\,10\,\% of the sky, to constrain cosmic inflation in the very early universe. The LAT will map $\sim$\,40\,\% of the sky at arcminute angular resolution to measure the number of relativistic species and the mass of neutrinos, improve our understanding of galaxy evolution, and constrain the properties of cosmic reionization.\cite{sofc19} 

\begin{figure}[b]
    \centerline{
    \includegraphics[width=1\linewidth]
{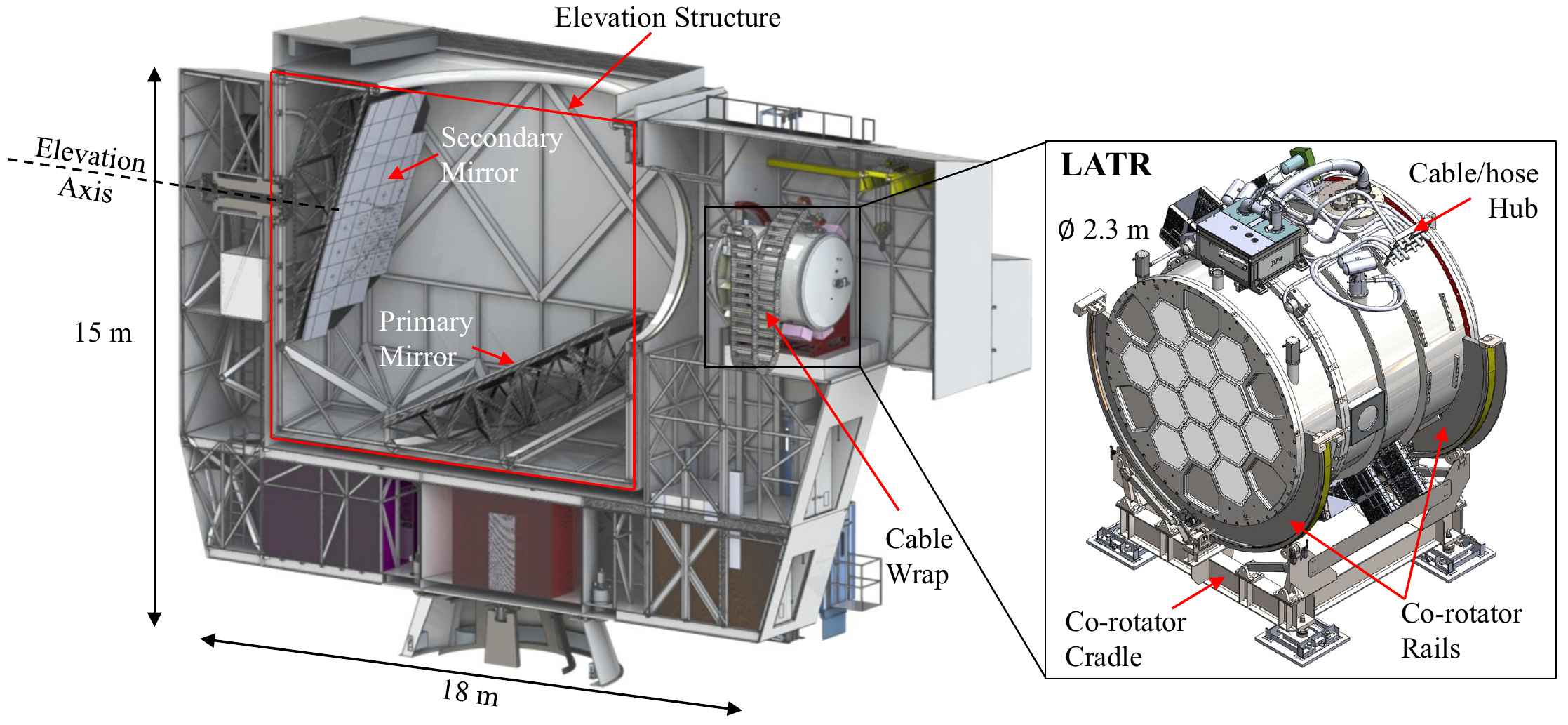}}
  \caption{The LAT and the LATR. The rendering on the left shows the cross section of the LAT with the LATR installed in the receiver cabin. The elevation structure is labeled with the 6\,m primary and secondary mirrors inside. Also labeled are the LATR and the flexible cable wrap. As the telescope changes observation elevation by rotating the elevation structure, the LATR co-rotates maintaining the same detector pointing and beam shapes. The co-rotator, as the interface between the LAT and the LATR, facilitates the rotation of the LATR. The inset view shows the co-rotator rails are bolted to the front and back flange of the cryostat. The four feet of the co-rotator cradle allows for fine adjustment when aligning the cryostat with the telescope mirrors.
    \label{fig:lat_n_latr}}
\end{figure}

The LAT adopted a coma-corrected, 6-meter aperture, crossed-Dragone optical design\cite{parshley/etal:2018} which feeds to a \diameter\,2.3\,m\, sub-100\,mK cryogenic receiver called the Large Aperture Telescope Receiver (LATR)\cite{zhu/etal:2018, orlowski/etal:2018, coppi/etal:2018}. Figure~\ref{fig:lat_n_latr} shows the rendering of the LAT and the LATR. This paper updates the previous publications\cite{zhu/etal:2018, orlowski/etal:2018, coppi/etal:2018} on the LATR design and describes the integration and validation of the instrument. An overview of the design and integration is shown in Section~\ref{sec:design}. Mechanical and cryogenic validation is presented in Section~\ref{sec:mech_validation},\,\ref{sec:cryo_validation}. The testing of the $\mu$Mux readout technology is presented in Section~\ref{sec:det_rd_validation}. We conclude in Section~\ref{sec:conslusion}.

\section{Design and Integration}
\label{sec:design}
The LATR is the cryogenic receiver of the SO LAT. During observation, the LATR co-rotates with the telescope elevation structure (Figure~\ref{fig:lat_n_latr}) at different pointing elevations\cite{gudmundsson/etal:2020, dicker/etal:2018}. The co-rotation maintains the relative detector pointing and beam shapes when the telescope observes at different elevations. The co-rotator, as shown in Figure~\ref{fig:lat_n_latr}, consists of a pair of co-rotating rails that are bolted to the front and back flange of the LATR, and a cradle that supports and rotates the rails. The design allows for a stable support structure for the LATR while allowing on-axis adjustments. The LATR hoses and electronic cables converge at the cable/hose hub on top of the LATR, before going into the cable wrap that comes from the telescope. The cable wrap ravels and unravels along the rotation of the LATR (Figure~\ref{fig:lat_n_latr}).

The LATR, coupled to the Large Aperture Telescope (LAT)\cite{parshley/etal:2018}, is designed to fill a focal plane diameter of 1.7\,m. The focal plane should be cooled to $<$\,100\,mK to provide operational temperature for $\sim$\,60,000 TES detectors\cite{irwin/hilton:2005}. In addition, the LATR should contain cold optics held at their designed locations with high precision. To achieve this goal, the LATR was divided into six temperature stages (300\,K, 80\,K, 40\,K, 4\,K, 1\,K, and 100\,mK) and 13 modularized optics tubes (see Figure~\ref{fig:latr_cut}). 

\subsection{LATR Design Overview}
\label{subsec:latr_design}

The 300\,K stage primarily serves as the cryostat's vacuum shell. In the front, a 6-cm-thick monolithic aluminum front plate with 13 openings, covered by plastic windows, resists atmospheric pressure while accepting signals from the openings. The structural strength of the vacuum chamber was carefully studied with finite-element analysis (FEA) to make sure the design holds the atmospheric pressure with the minimal amount of material. Anti-reflective (AR) coated ultra-high-molecular-weight polyethylene (UHMWPE) windows and double-sided IR-blocking (DSIR) filters are installed on each of the 13 openings. The windows are made from 3.2\,mm UHMWPE sheets and AR-coated with porous teflon sheets on both sides. The windows hold the vacuum and provide high in-band optical throughput. The DSIR filters reflect away thermal loading from infrared radiation before it enters the cryogenic stages.

\begin{figure}[t]
    \centering
    \includegraphics[width=1\textwidth]{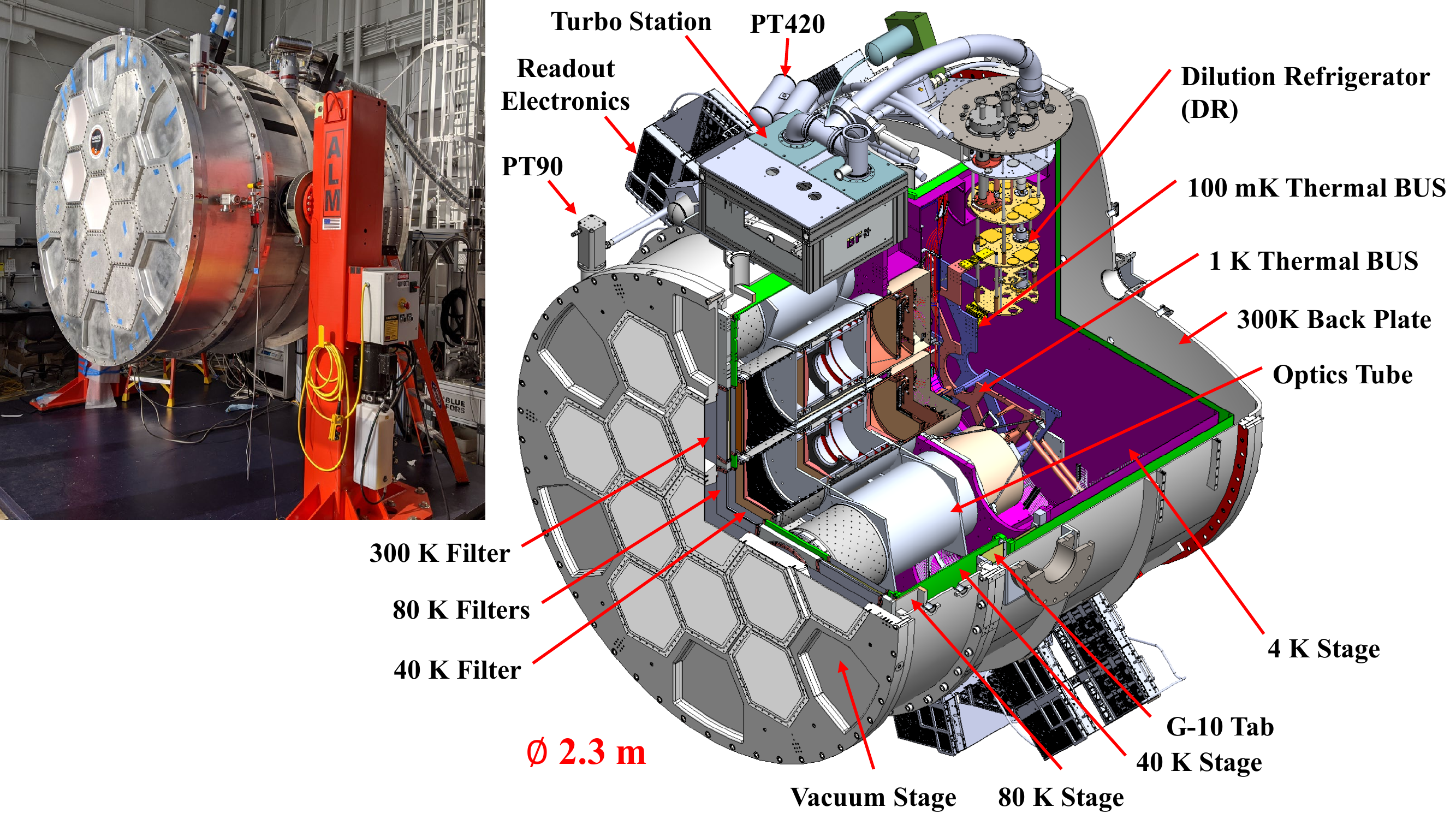}
    \caption{A cross-section view of the LATR. The rendering shows the 300\,K vacuum stage, 80\,K stage (dark grey), 40\,K stage (green), and 4\,K stage (purple). Section~\ref{sec:design} describes the design of each stage in detail. One of the 300-40\,K G-10 tabs (more details in the text) is visible and labeled. Inside the 4\,K cavity, the 1\,K and 100\,mK buses distribute the cooling power from the dilution refrigerator (DR) to individual optics tubes (OT). Infrared filters on the cryostat are shown at 300\,K, 80\,K, and 40\,K stages in front of the OTs. The OTs contain optical elements at $\le$4\,K and the detector arrays. The DR and its turbo station are shown in the upper part of the LATR. One PT420 and one PT90 pulse-tube coolers along with the warm readout electronics\cite{sathyanarayana/etal:2020} are also annotated around the LATR. A photo of the LATR is added on the top left during its test in the highbay at the University of Pennsylvania.}
    \label{fig:latr_cut}
\end{figure}

After the 300\,K front plate, a monolithic aluminum\footnote{Aluminum alloy 1100-H14 is used here for its high thermal conductivity.} 80\,K plate was designed, with 13 matching openings. Each 80\,K opening is equipped with another IR-blocking filter and an absorptive alumina filter. The IR-blocking filter reflects away the thermal loading similar to the ones at 300\,K, and the alumina filters absorb thermal radiation that passes through the reflective filters and conducts the heat away. The alumina filters are also designed to be wedge shaped at off-center locations to make the chief ray of the central field of each opening parallel to the axis of symmetry\cite{gudmundsson/etal:2020, dicker/etal:2018}. Considering the high dielectric constant, the alumina filters are anti-reflective coated on both sides\cite{jeong/etal:2019}. The 80\,K plate constitutes most of the 80\,K stage which is cooled down by two single-stage pulse-tube cryogenic coolers (PT90) from Cryomech\footnote{\label{note:cryomech}Cryomech, website: \href{https://www.cryomech.com/}{https://www.cryomech.com/}}. The two pulse-tube coolers provide 180\,W cooling power at 80\,K, which is effectively distributed across the 2\,m diameter by a design detailed in Ref.~\citenum{coppi/etal:2018, zhu/etal:2018, orlowski/etal:2018}. The entire 80\,K stage is covered with 30 layers of multi-layer insulation (MLI)\footnote{\label{note:ruag}RUAG Space. Website: \href{https://www.ruag.com/en/products-services/space/spacecraft/multi-layer-insulation}{https://www.ruag.com/en/products-services/space/spacecraft/multi-layer-insulation}} to reduce radiative thermal loading.

Behind the 80\,K stage, another single-piece aluminum plate with a diameter of 2.08\,m is designed to operate at 40\,K. This 40\,K plate is the front of the 40\,K stage, which extends to the back of the cryostat (color-coded green in Figure~\ref{fig:latr_cut}). Each of the openings is equipped with another IR-blocking filter to further reflect away infrared thermal radiation. The 40\,K stage encloses the colder stages and is primarily cooled down by another two pulse-tube coolers (PT420) from Cryomech, providing $\sim$\,100\,W total cooling power at 40\,K. The two PT420s are mounted in the mid-section of the 40\,K cylinder to effectively deliver the cooling power across the 40\,K stage. Given the $>$\,2\,m span in length of the 40\,K stage, high-purity 5N aluminum bars are added on the 40\,K shell to effectively distribute the cooling power to the two ends. Considering the large surface area of the 40\,K stage, 30 layers of MLI are wrapped around to reduce thermal loading from radiation.

Starting from the 4\,K stage and below, the cryogenic components, including filters, lenses, and detector arrays, are packaged in 13 individual modules, called optics tubes (OTs). The modularizad design facilitates the manufacture and construction of the components at $<$\,4\,K. More importantly, it provides strict controls on the relative positions of the optical elements. The 13 optics tubes are installed on a 2.54\,cm thick monolithic aluminum\footnote{Aluminum alloy 1100-H14 is used here for its high thermal conductivity.} 4\,K plate to meet the thermal and mechanical requirements. The 4\,K stage starts with the 4\,K plate and extends to the back of the cryostat (shown in purple in Figure~\ref{fig:latr_cut}). The 4\,K stage, along with the 4\,K part of the 13 optics tubes, is primarily cooled down by the second stage of the two PT420s, providing $\sim$\,4\,W total cooling power at 4\,K. The exterior of the 4\,K stage was wrapped with 10-layer MLI, a lower requirement since it is already within the 40\,K stage.

\begin{figure}
  \centering
    \includegraphics[width=\linewidth]{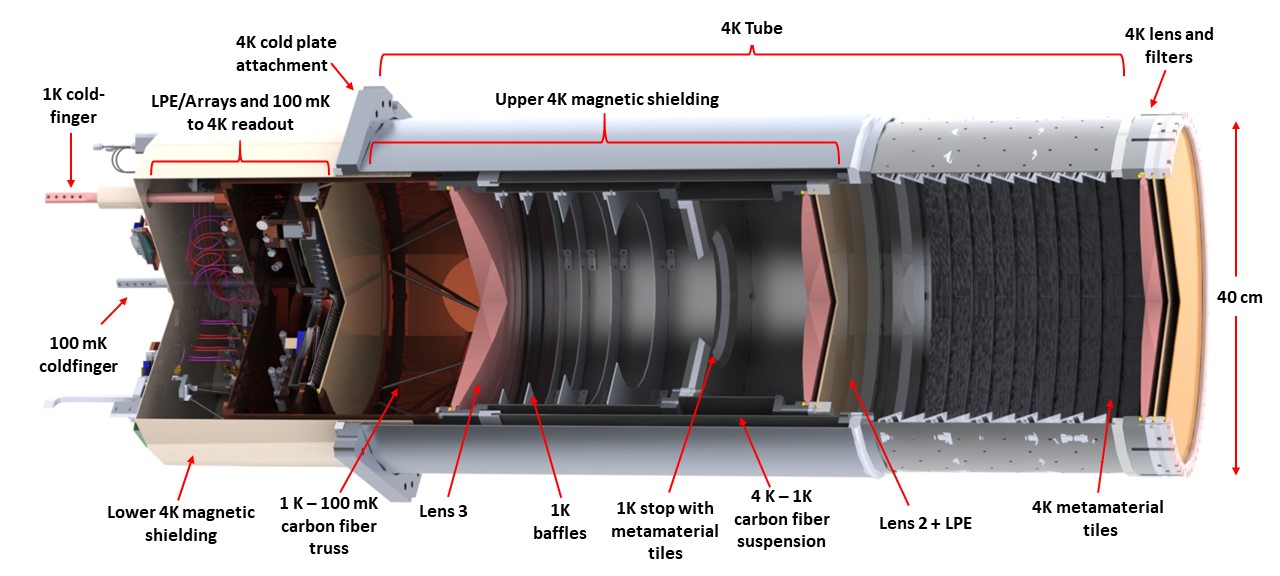}
  \caption{This rendering shows major parts of one optics tube (OT). Light rays enter from the right through the 40-cm-diameter 4\,K lens and filters, refracted by lens\,2 before truncated by the Lyot stop. Then the light rays are focused by lens\,3 before being detected by the detector arrays. The 4\,K metamaterial tile section, the metamaterial absorber covering Lyost stop, and the 1\,K baffle section are designed to terminate stray light. Other mechanical, cryogenic, and RF components are also annotated in the rendering. The RF components at the back of the OT is further discussed in Section~\ref{sec:det_rd_validation}.}
  \label{fig:OT_cutaway}
\end{figure}

A dilution refrigerator (DR)\footnote{Bluefors LD400, website: \href{https://bluefors.com/products/ld-dilution-refrigerator/}{https://bluefors.com/products/ld-dilution-refrigerator/}}, capable of providing $>$\,400\,$\mu$W cooling power at 100\,mK, is integrated into the back of the LATR. The DR provides cooling power for 1\,K and 100\,mK stages in the 13 optics tubes. To effectively distribute the cooling power to the 13 optics tubes on a diameter of 2\,m, two thermal buses\footnote{The use of bus is analogous to that in computers, but conducting heat in this context.} were designed for 1\,K and 100\,mK respectively (as shown in Figure~\ref{fig:latr_cut}). The two thermal buses are hexagon web structures (Figure~\ref{fig:latr_cut} and \ref{fig:4k_rf}) made of oxygen-free-high-conductivity copper, gold-plated to minimize radiative thermal loading. The two thermal buses deliver cooling power right to the back of the 13 OTs before cold fingers (shown in Figure~\ref{fig:OT_cutaway}) connect the buses to the corresponding stages in the OTs.

The OTs are the core of the LATR, around which the entire cryostat is designed. Each OT is roughly 40\,cm in diameter and 130\,cm long and is designed to be installed and removed independently. Each OT uses three anti-reflective coated silicon lenses to re-image the telescope focal plane onto three detector arrays\cite{gudmundsson/etal:2020, dicker/etal:2018}. A rendering of the optics tube design is shown in Figure~\ref{fig:OT_cutaway}. Each OT is mechanically supported by the 4\,K plate attachment flange and both the upper and lower part of the OT launch from there. The upper OT at 4\,K extends from the 4\,K plate to the back of the 40\,K plate mentioned before, where another IR-blocking filter, a low-pass edge (LPE) filter, and the first silicon lens are installed as the 4\,K optical elements. The LPE filter, together with the other two to be introduced, is a metal mesh reflective filter designed to gradually reduce out-of-band loading at successive thermal stages\cite{ade/etal:2006}. Absorbing tiles\cite{xu/etal:2020b} are installed on the wall at the 4\,K section right after the 4\,K optics. Optical simulation reported by Ref.~\citenum{gudmundsson/etal:2020} shows that effectively blackening this part increases the telescope sensitivity by $\sim$\,40\,\%. 

The 1\,K stage in the optics tube is suspended on a thermally-isolated carbon-fiber tube from the 4\,K stage (see Figure~\ref{fig:OT_cutaway}). 
The 1\,K stage contains another LPE filter, the second silicon lens, the absorber-covered\cite{xu/etal:2020b} Lyot stop, ring baffles, and the third silicon lens. The Lyot stop and the following ring baffles effectively truncate the beam, reject stray light, and terminate the stray light at 1\,K temperature. All the LPE filters and lenses are mounted with spring-loaded mechanism for cryogenic survivability and optical alignment purposes. Finally, the 100\,mK stage is launched from the back of the 1\,K stage with carbon-fiber trusses (see Figure~\ref{fig:OT_cutaway}). Another LPE filter is mounted on the 100\,mK stage followed by the three detector arrays. 
Behind the detector array, the $\mu$Mux\cite{sathyanarayana/etal:2020} readout technology necessitates a series of DC and RF feedthroughs out the back end of the OTs (see Section~\ref{sec:det_rd_validation}). Coldfingers at 1\,K and 100\,mK launch from the OT stages to thermally connect to the thermal buses in the cryostat (see Figure~\ref{fig:latr_cut}). Magnetic shielding is designed around the back and extends to the front part of the OTs. Combined with the magnetic shielding on the detector arrays, simulation shows the overall magnetic shielding strategy delivers sufficient attenuation of environmental magnetic fields.\cite{vavagiakis/etal:2020}

\subsection{Alignment Design Strategy}
\label{subsec:mech_align_design}

As an optical instrument, optical elements in the LATR should be aligned. Given the size and weight of each temperature stage, the internal mechanical support structure should balance the competing requirements of strength, to minimize movement or deformation as the telescope observes, with thermal isolation, to allow for cryogenic operations. Additionally, one should allow for the thermal contraction as the structure is cooled to the operating cryogenic temperatures. The contraction is significant ($\sim$\,2\,cm) over the $>$\,2\,m LATR diameter and varies between materials and the different operating temperatures.
The mechanical design should overcome all the challenges while maintaining millimeter-level alignment. The solution was a combination of precisely-fabricated G-10 tabs\cite{gudmundsson/etal:2015} and modularized OTs\cite{thornton/etal:2016}. 

\begin{figure}
    \centering
    \includegraphics[width=0.5\textwidth]{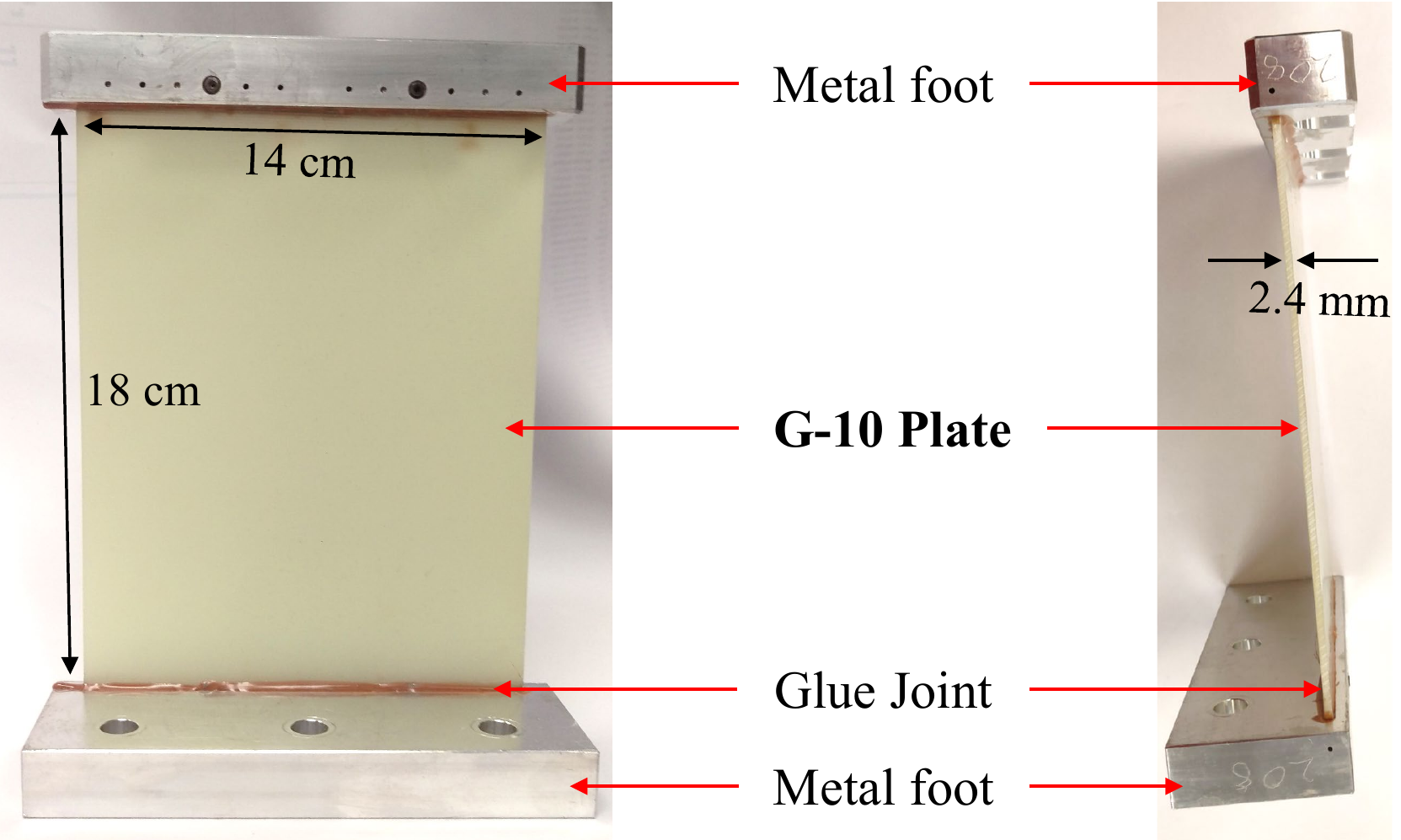}
    \caption{One 300-80\,K G-10 tab is shown from two perspectives. In addition to the G-10 plate, two metal feet are glued on the two ends as mechanical interfaces. Armstrong A-12 was chosen as the glue considering its mechanical and cryogenic performance. Each G-10 tab is serial numbered as shown in the side-view photo on the right.}
    \label{fig:g10_tab}
\end{figure}

G-10 is a thermally-isolating material with great strength. Separating them into evenly-spaced individual tabs further reduced the thermal conductivity while allowing the mechanical compliance required by thermal contraction. In the LATR, the 80\,K stage is supported by 12 G-10 tabs from the front of the 300\,K stage; the 40\,K stage is supported by 24 G-10 tabs from the middle of the vacuum shell; and the 4\,K stage (and below) is supported by another 24 G-10 tabs from the 40\,K stage. The tabs are all 2.4\,mm thick with slightly different length and width\footnote{The G-10 material meets NIST G-10 CR process specification and conforms to MIL-I-247682 Type GEE/CR.} (Table~\ref{tab:g10_tab}). One 300-40\,K G-10 tab is annotated in Figure~\ref{fig:latr_cut} and photos of one 300-80\,K G-10 tab are shown in Figure~\ref{fig:g10_tab}.

\begin{table}[h]
\begin{center}       
\caption{G-10 tabs in the LATR. All G-10 tabs are 2.4\,mm thick with different widths and lengths shown in the table. Also shown are the number of the tabs for each stage.}
\label{tab:g10_tab}
\begin{tabular}{c c c c} 
\hline
\hline
Stage & Width (mm) & Length (mm) & Number of Tabs\\

\hline
300-80\,K  & 140 & 180 & 12 \\
300-40\,K & 160 & 150 & 24  \\
40-4\,K & 150 & 194.5 & 24 \\

\hline \\
\end{tabular}
\vspace{-0.7cm}
\end{center}
\end{table}

Below the 4\,K stage, the optical elements and the detector arrays are assembled in the modularized OTs. This small-scale OT assembly assures high precision alignment internally. As long as we precisely position the OTs on the 4\,K plate, the OT internal components are all aligned. Therefore, alignment of the 4\,K plate (relative to other higher temperature stages) and the internal OT alignment together deliver the alignment of the entire optical chain. 

\section{MECHANICAL VALIDATION}
\label{sec:mech_validation}
The LATR mechanical validation aims to verify that the mechanical design holds atmospheric pressure and supports the weight of the internal structures. In addition, it also verifies that optical elements in the LATR are aligned within optical requirements.

Over its 27\,m$^2$ exterior surface, the atmosphere exerts $\sim300$-metric-ton forces on the LATR vacuum shell. The LATR has been pumped down for over 20 times and has stayed under vacuum for months, validating the strength of the vacuum shell. Internal temperature stages, held by G-10 tabs and carbon fiber structures, have also been tested for 12 thermal cycles under various loads, validating the strength of the internal supporting structures.

Beyond the mechanical strength, optical simulations set constraints on the alignment of the optical elements\cite{gudmundsson/etal:2020, dicker/etal:2018}. We validated the alignment of the optical chain in two steps:
\begin{enumerate}
    \item Cryostat-level: validating the alignment of the single-piece plates at 300\,K, 80\,K, 40\,K, and 4\,K stage, especially the 4\,K stage since it is the anchor point for all the OTs.
    \item OT-level: validating the alignment of the lenses, the Lyot stop, and the detector arrays within OTs, with respect to their mounting flanges to the 4\,K plate.
\end{enumerate}

All the measurements are conducted at room temperature. We do not expect the achieved alignment to change significantly at cryogenic temperatures. However, cryogenic measurements are planned and will be reported in future publications.

\subsection{Cryostat Alignment Validation}
\label{subsec:cryostat_alignment}

\begin{figure}
    \centering
    \includegraphics[width=\linewidth]{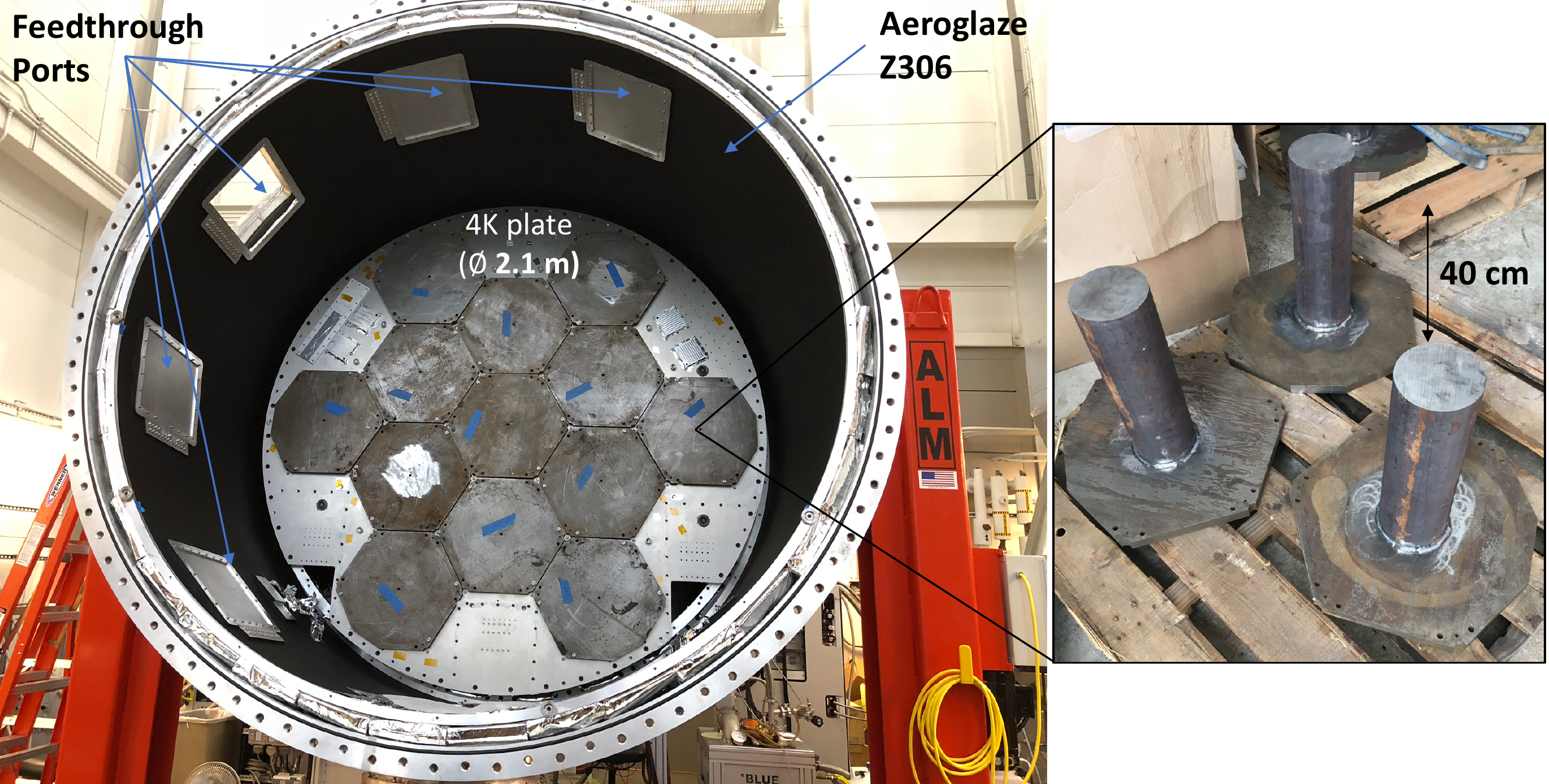}
    \caption{LATR metrology setup with dummy weights. On the left, the back of the LATR is shown with the back plates off. The 4\,K plate (with 13 dummy weights installed) is exposed during the 13-OT load weight test. Five feedthrough ports are also seen on the cylindrical shield. The interior wall of the 40\,K shell is coated with Aeroglaze Z306 to absorb 300\,K photons leaked into the 40\,K stage. A close-up photo of the dummy weights is shown on the right. The dummy weights simulate not only the weight of each optics tube, but also the center of mass, achieved by a steel plate welded with a steel bar.}
    \label{fig:weight_test}
\end{figure}

Incoming signal passes through openings on the single-piece 300\,K, 80\,K, and 40\,K plates before it enters the modulairzed OTs. Openings on these plates are normally equipped with planar windows and filters, without active optical elements.\footnote{The 80\,K plate openings have wedge-shaped alumina filters. However, simulations show that alignment requirements on these filters are more forgiving than those of the 80\,K plate.\cite{gudmundsson/etal:2020, dicker/etal:2018}} Therefore, the positioning requirement for these plates is that each of the openings should be concentric to the extent that the geometric beams are not clipped. Openings on the plates are designed with at least 1\,cm margin, meaning the plates should be precisely positioned within 1\,cm.

The more demanding positioning requirement comes with the 4\,K plate, which mounts the 13 OTs, full of active optical elements. Optical simulation\cite{gudmundsson/etal:2020, dicker/etal:2018} shows that position of the OTs should be maintained within $\pm6$\,mm along the LATR's optical axis and $\pm4$\,mm perpendicular to the LATR's optical axis. In addition, the OTs should not exceed $\pm 0.8 ^{\circ}$ in tilt with respect to each other. These constraints need to be met when the cryostat is initially deployed with a small number of OTs as well as when it is fully loaded with 13 OTs. The requirement should also be met when the LATR is rotated during observation.

To ensure these tight tolerances are met, we utilized a state-of-the-art metrology system from FARO Technologies.\footnote{FARO website: \url{https://www.faro.com/}} The FARO Vantage Laser Tracker  was used to measure the locations of all relevant plate surfaces both internal to, and external to, the cryostat. The FARO Vantage Laser Tracker has an accuracy of 16\,$\mu$m and a single point repeatability of 8\,$\mu$m at 1.6\,m. The Vantage Laser Tracker was used to verify the overall dimensions of the cryostat, the 3D locations of the single-piece plates, as well as the precise locations of the optics tube mounting flanges on the 4\,K plate. During the factory acceptance tests, we performed comprehensive measurements around the cryostat and verified that the cryostat, especially the four single-piece plates, was built within the required tolerances. Since the 4\,K plate position can be affected by the additional weight from 13 optics tubes, we subsequently performed measurements of the 4\,K stage with 13 mass dummies simulating the weight and the torque that they will exert on the 4\,K plate (Figure~\ref{fig:weight_test}). Some details of the measurements are presented in Table~\ref{tab:4k_metrology}.

\begin{table}[h]
\begin{center}       
\caption{LATR 4\,K plate metrology results. This table shows some measured positions of the 4\,K plate under two conditions: no-load and 13-OT loaded. Under both conditions, the plate location is within the required tolerances\cite{gudmundsson/etal:2020, dicker/etal:2018}. We use the center of the 4\,K plate as the reference. All deviation values are absolute values from the designed location. At these deviation levels, we expect the tilting requirements are met; more detailed analysis on tilting will be presented in future publications. Axes are defined as (in Figure~\ref{fig:latr_cut}): the x-axis parallel to the ground, the y-axis perpendicular to the ground, and the z-axis along the cryostat optical axis.}

\label{tab:4k_metrology}
\begin{tabular}{c c c c} 
\hline
\hline
Axis & Tolerance & Deviation from Design (no-load) & Deviation from Design (13-OT load)  \\
\hline
X-axis & 3\,mm & 0.49 $\pm$ 0.03\,mm & 0.31 $\pm$ 0.02\,mm\\
Y-axis & 3\,mm & 2.04 $\pm$ 0.03\,mm & 2.24 $\pm$ 0.02\,mm\\
Z-axis & 3\,mm & 2.03 $\pm$ 0.03\,mm & 2.43 $\pm$ 0.02\,mm\\
\hline \\
\end{tabular}
\vspace{-0.3cm}
\end{center}
\end{table}

Another main concern about the 300\,K vacuum shell is the deformation of the front plate with 13 openings covered by plastic windows. Finite-element analysis (FEA) on the complicated front plate predicts a bow-in of 18\,mm under atmosphere, which is verified by the measured bow-in as 17\,mm.

\subsection{Optics Tube Alignment Validation}
\label{subsec:ot_alignment}
After validating the OT mounting locations on the 4\,K plate, optical elements within the OTs should be correctly positioned with respect to the mounting flange. All three lenses within the OTs should maintain a position of $\pm 2$\,mm along the LATR's optical axis and $\pm 2$ \,mm perpendicular to the LATR's optical axis. Each lens should stay within $\pm 0.4^{\circ}$ with respect to the plane perpendicular to the optical path.

Due to the size and intricate design of the OTs, we utilized the FARO Edge ScanArm to obtain precise cold optical component and detector array locations. The metrology is performed on tubes that are not installed in the LATR to allow for easier access to hard-to-reach locations within the tubes. The FARO Edge ScanArm has an accuracy of 34\,$\mu$m and a repeatability of 24\,$\mu$m at 1.8\,m. Each individual mechanical component of every OT is measured and recorded. Partial and full sub-assemblies of the OT are also measured, thus allowing us to understand the position of, parallelism between, distance between, and coaxiality of all components within each OT. Tight constraints on the locations of the focal plane array and lenses creates a need for extensive documentation of all dimensions of OT components. From these measurements, we have been able to confirm that all the optical elements of the optics tube, the lenses, the Lyot stop, and the detector arrays, are mounted within the required mechanical tolerances\cite{gudmundsson/etal:2020, dicker/etal:2018}. Some metrology results within one OT are presented in Table~\ref{tab:OT_metrology}. 

\begin{table}
\begin{center}       
\caption{Metrology results demonstrate that positions of optical elements at room temperature lie within the required tolerances. This table lists the measured absolute deviations from the design positions and allowed tolerances along the z-axis\cite{gudmundsson/etal:2020, dicker/etal:2018} for one OT. The z-axis travels along the symmetrical axis of one optics tube. A more detailed analysis will be presented in future publications.}
\label{tab:OT_metrology}
\begin{tabular}{c c c} 
\hline
\hline
Optical Element & Z-axis Tolerance & Deviation from Design  \\
\hline
Lens 1 & 2.0\,mm & 0.47 $\pm$ 0.02\,mm\\
Lens 2 & 2.0\,mm & 0.19 $\pm$ 0.02\,mm\\
Lens 3 & 2.0\,mm & 0.50 $\pm$ 0.02\,mm\\
Lyot Stop & 2.0\,mm & 0.22 $\pm$ 0.02\,mm\\
Detector Arrays & 2.5\,mm & 0.53 $\pm$ 0.02\,mm\\
\hline \\
\end{tabular}
\vspace{-0.7cm}
\end{center}
\end{table}

\section{CRYOGENIC VALIDATION}
\label{sec:cryo_validation}
The goal of the cryogenic validation is to verify that all parts of LATR cool down to required temperatures within expected time. Since not all OTs are available now, we aim to measure the loading from the available configuration and project the performance with 13 OTs, and verify that the LATR meets the cryogenic requirements in the full configuration.

\subsection{Cryogenic Validation Strategy}
\label{subsec:cryo_val_strategy}

Knowing that each pulse-tube cooler unit shows different cooling performance even from the same model, we calibrated the cryogenic coolers (two PT90s, two PT420s, and the Bluefors DR) individually. The calibration results provide specific cooling performance for each unit. This information enables us to accurately calculate the thermal loading on each temperature stage.

The cryogenic validation of the LATR was conducted incrementally, starting with the simplest configuration and gradually adding more factors in. We first conducted a 4\,K dark test, with two PT420s and two PT90s only. Then we tested the system after installing the DR and the 1\,K/100\,mK thermal buses. Thermal loadings on each stage were then calculated from the calibrated cryogenic cooler performance. With all the openings on the 300\,K plate covered with metal plates, these two dark tests provided a baseline result without optical loading.

After the dark tests, we started to install the plastic windows and filters at 300\,K, 80\,K, and 40\,K stages. We first tested the configuration with two windows+filter sets installed (2-window configuration) followed by another configuration with three windows+filter sets installed (3-window configuration). The 2-window configuration does not have OTs installed and the 4\,K plate openings were covered with metal plates to reject optical loading to $\le$\,4\,K stages. The 3-window configuration has one OT and one universal readout harness (URH) installed (see Section~\ref{subsec:umux_latr} for more details on the URH). Given the overall progress of the project, only three window+filter sets were initially fabricated to provide enough fidelity from the tests.

\subsection{Cryogenic Validation Results}
\label{subsec:cryo_val_result}

Among all the $\ge$4\,K stages, the 80\,K stage is the most sensitive to radiation loading because of the absorptive alumina filters. Measurements of the 80\,K stage thermal loading under different configurations are summarized in the top part of Table~\ref{tab:latr_thermal}. The baseline loading from the dark test is $\sim22$\,W with each window+filter set adding another $\sim$\,7\,W radiation loading on the 80\,K stage. If we project the results to 13 windows, the anticipated loading will be $\sim$\,113\,W. Given the calibrated PT90 cooling performance, we deduced that the 80\,K stage will stay at around 65\,K, safely below the designed 80\,K requirement. The temperature gradient across the 2.1\,m diameter plate is measured at $\sim$\,3\,K in the 3-window configuration. Receiving roughly three times the loading with 13 optics tubes, the gradient is expected to increase to $\sim9$\,K.

We observed a difference between the measured loading and the predicted loading on the 80\,K stage. Predictions from the thermal model do not include any imperfections in the MLI installation. Considering the size of the 2.1-m-diameter 80\,K plate and the complicated geometry of the G-10 tabs, it is unsurprising to observe extra loading due to imperfect shielding of the surfaces. Beyond the dark test, we measured $\sim$\,7\,W per filter set instead of the predicted $\sim$\,3\,W. We are still investigating the mismatch which is likely a result from the higher-than-expected infrared transmission of the IR-blocking filters.

\begin{table}
\caption{This table shows base temperatures and thermal loading for the 80\,K, 40\,K, and 4\,K stage under different configurations. The temperature range is measured from six thermometers evenly distributed on the 80\,K and 40\,K plates and five for the 4\,K plate.  In the 3-window configuration, one OT and one URH were also installed in addition to 3 windows+filter sets.} 

\label{tab:latr_thermal}
\begin{center}       
\begin{tabular}{c c c} 
\hline
\hline
Configuration & Stage Temperature & Measured Power\\
\hline
\multicolumn{3}{c}{\textbf{80\,K Stage}}\\
\hline
Dark & 37\,--\,39\,K & $22_{-1}^{+1}$\,W\\
2-window & 40\,--\,43\,K & $35_{-1}^{+1}$\,W\\
3-window & 44\,--\,47\,K & $42_{-1}^{+1}$\,W\\
\hline 
\multicolumn{3}{c}{\textbf{40\,K Stage}}\\
\hline
Dark & 44\,--\,47\,K & $33_{-1}^{+1}$\,W \\
2-window & 44\,--\,47\,K & $<33$\,W \\
3-window & 44\,--\,48\,K & $<34$\,W\\
\hline
\multicolumn{3}{c}{\textbf{4\,K Stage}}\\
\hline
Dark & 3.5\,--\,5.2\,K & $0.8_{-0.1}^{+0.1}$\,W  \\
2-window & 3.6\,--\,5.0\,K & $0.8_{-0.1}^{+0.2}$\,W  \\
3-window & 3.8\,--\,5.2\,K & $1.3_{-0.2}^{+0.2}$\,W\\
\hline
\vspace{-1cm}
\end{tabular}
\end{center}
\end{table}

During the no-DR dark test, the 40\,K plate stayed at 44\,--\,47\,K (measured at six locations on the plate) with an estimated loading of $33\pm1$\,W; the 4\,K plate stayed at 3.5\,--\,5.2\,K (measured at five locations on the plate) with an estimated loading of $0.8\pm0.1$\,W, as shown in the middle and bottom part of Table~\ref{tab:latr_thermal}. Estimating the thermal loading, especially on the 4\,K stage, is challenging since the amount of thermal loading is on the lower end of anticipation where the pulse-tube cooler calibration data are sparse. At this low-loading range, we estimate the measurement uncertainties as $\sim$\,1\,W on the 40\,K stage and $\sim$\,0.1\,W on the 4\,K stage. To reassure the measurement accuracy of the 4\,K stage measurement, we installed seven heaters (evenly distributed) on the 4\,K plate to dissipate additional power and monitor the temperature change. From those data, we calculated the thermal conductance of the thermal links connecting the pulse-tube 4\,K cold heads and the 4\,K plate and then extrapolated to calculate the intrinsic loading of the 4\,K stage. This independent measurement gives consistent results with the ones from the calibrated cooler performance.

After the DR was installed (starting from the 2-window configuration in Table~\ref{tab:latr_thermal}), estimating thermal loading on the 40\,K and 4\,K stage became more difficult because the PT420 on the DR also contributed to the cooling of the 40\,K and 4\,K stages. We did not calibrate the PT420 in the DR since it is deeply integrated in the DR system. Furthermore, the thermal linking between the DR 40\,K/4\,K stage to the cryostat 40\,K/4\,K stage is difficult to quantify. We measured the DR 40\,K and 4\,K stage temperatures and compared to the values previously measured in its stand-alone cryostat as an adiabatic reference.

Interestingly, the DR 40\,K temperature stage decreased from its stand-alone measurement to the measurements in the LATR. This means that a fraction of the 40\,K stage cooling power from the main cryostat goes to the DR 40\,K stage. Therefore, we report the 40\,K thermal loading from the two calibrated PT420s as upper limits in Table~\ref{tab:latr_thermal}. The DR 4\,K temperature did not change in the 2-window configuration and rose by $\sim$0.2\,K in the 3-window configuration, mainly because of the addition of the OT and the URH. Using the average of the calibrated cooling performance from the two PT420s, we estimated the additional cooling power from the observed temperature change. Together with the power from the two calibrated PT420s, the thermal loading measurements are reported in Table~\ref{tab:latr_thermal}. The measured 4\,K loading is higher than expectation with relatively large uncertainties. We are making additional measurements to further investigate the issue. The results will be reported in future publications. Note that the available cooling power from the two PT420s ($>$\,110\,W at 40\,K; $>$\,4\,W at 4\,K) is around three times more than the estimated thermal loading from the 3-window configuration. 

The temperature gradient across the 2.06\,m diameter 40\,K filter plate is $\le$\,4\,K in the 3-window configuration. Note that the 40\,K plate is the farthest away from the 40\,K pulse-tube cold heads (Figure~\ref{fig:latr_cut}). From the dark to the 3-window configuration, we did not measure significant changes on the 40\,K plate, in terms of both thermal loading and thermal gradient. This result is consistent with the simulation prediction because only reflective filters are installed on the 40\,K plate.  
The measured temperatures and the small thermal gradient proved that the pulse-tube cooling power was efficiently distributed around the 2\,m long 40\,K stage to maintain the entire stage below 50\,K. The thermal gradient across the 4\,K plate is around 1.7\,K during the dark configuration, while reduced to 1.4\,K after installing the DR. The addition of the windows did not significantly change the loading on the 4\,K which is consistent with the simulation. In the 3-window configuration, the temperature on the 4\,K shell is $\sim$\,5\,K. This measurement validates that the entire 4\,K stage was efficiently cooled down with 3 window+filter sets, one OT, and one URH installed.

Enclosed within the 4\,K stage cavity, the 100\,mK thermal bus cooled to $<50$\,mK with the 1\,K thermal BUS maintained at $\sim$\,1\,K without optics tubes installed. For tests without the optics tube, heaters were installed on both the 100\,mK and 1\,K thermal BUS to simulate expected thermal loads. With these two heaters, we were able to map out the load curve on the two stages which later informed us of the loading from the addition of one OT. We also dissipated the anticipated loading from 13 optics tubes on the two stages when the 100\,mK thermal bus stayed below 100\,mK with $<$10\,mK thermal gradient across. 

\begin{table}
\begin{center}       
\caption{LATR 100\,mK stage cryogenic performance. Measurements from 1-OT configuration is shown together with the projection for the fully-equipped 13-OT configuration. Upper limits on the projected temperatures are presented given the uncertainties of the measurement. The projected OT temperature shows the highest number among the 13 OTs.}

\label{tab:100mk_cryo}
\begin{tabular}{c c c c c c c} 
\hline
\hline
\multirow{2}{*}{Configuration} & \multicolumn{3}{|c|}{Loading} & \multicolumn{3}{c}{Temperature}\\\cline{2-7}
& \multicolumn{1}{|c}{Optics Tube} & BUS & \multicolumn{1}{c|}{Total} & DR & BUS & Optics Tube  \\
\hline
1-OT & $3.5^{+1.5}_{-1.0}$\,$\mu$W & $20^{+5}_{-10}$\,$\mu$W & $24^{+5}_{-10}$\,$\mu$W & 32\,mK & $50$\,mK & 56\,mK\\

13-OT* & $46^{+20}_{-13}\,\mu$W & $20^{+5}_{-10}$\,$\mu$W & $66^{+21}_{-16}$\,$\mu$W & $\le50$\,mK & $\le75$\,mK & $\le95$\,mK\\
\hline
\end{tabular}
\vspace{-0.3cm}
\end{center}
\end{table}

\begin{figure}[b]
    \centering
    \includegraphics[width=\textwidth]{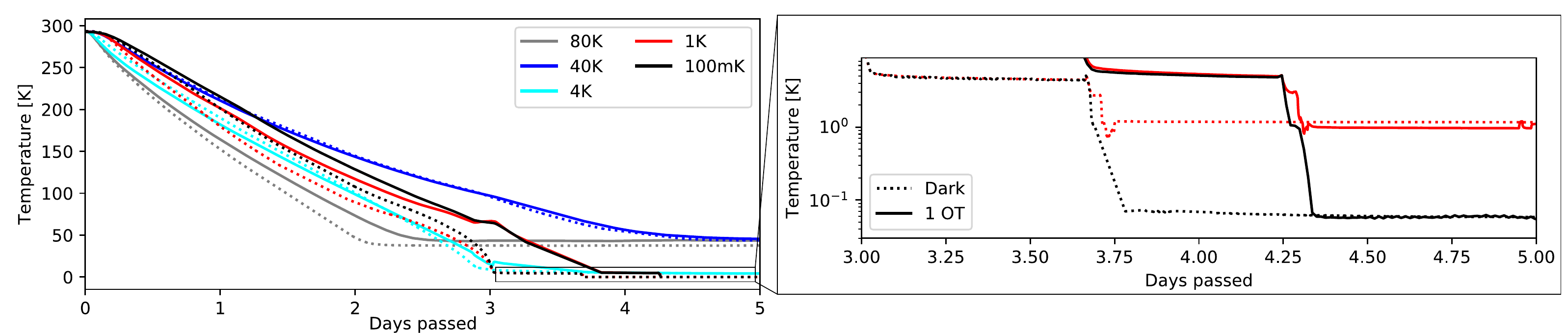}
    \caption{Cool-down curves from two test runs: the dark configuration (with the DR) and the 3-window (+ one OT + one URH) configuration. The left panel shows the cool-down curves of different temperature stages. Although multiple thermometers are installed on each temperature stage, we are only plotting a representative one for simplicity. Temperatures measured during the dark configuration are represented by dotted lines, and measurements from the 3-window configuration are displayed with solid lines. Under these configurations, all the stages cooled down to base temperatures within 5 days. The right panel is a zoom-in of the 1\,K and 100\,mK stages in a logarithmic scale. These two stages reached base temperatures within hours after the DR was turned on.}
    \label{fig:cool-down_curve}
\end{figure}

After installing one OT (along with all its RF readout components), its thermal loading on the 100\,mK stage was measured. We used the calibrated DR performance to calculate that the OT dissipates $3.5^{+1.5}_{-1.0}\,\mu$W\footnote{This result was later confirmed by an independent OT test cryostat as $4.2\pm0.2\,\mu$W.\cite{harrington/etal:2020}} thermal loading at 100\,mK and the 100\,mK thermal bus receives $20^{+5}_{-10}\,\mu$W parasitic thermal loading from cabling, mechanical supports, etc. Loading from 1\,K stage is much less than the DR capacity therefore is not discussed in detail. If we use the 1-OT measurement as a reference and extrapolate to 13 OTs, the expected loading is $66^{+21}_{-16}\,\mu$W, raising the DR to $\le50$\,mK. Thermal gradient from the DR 100\,mK stage to each OT 100\,mK stage is calculated with the calibrated properties of the thermal interfaces. The results show that the highest 100\,mK stage temperature among the 13 OTs are projected to be $\le$\,95\,mK, lower than the 100\,mK requirement. More details of 100\,mK stage results are presented in Table~\ref{tab:100mk_cryo}.

\subsection{Cool-down Time Results}
\label{subsec:cool-down_time}

\begin{figure}
    \centering
    \includegraphics[width=\linewidth]{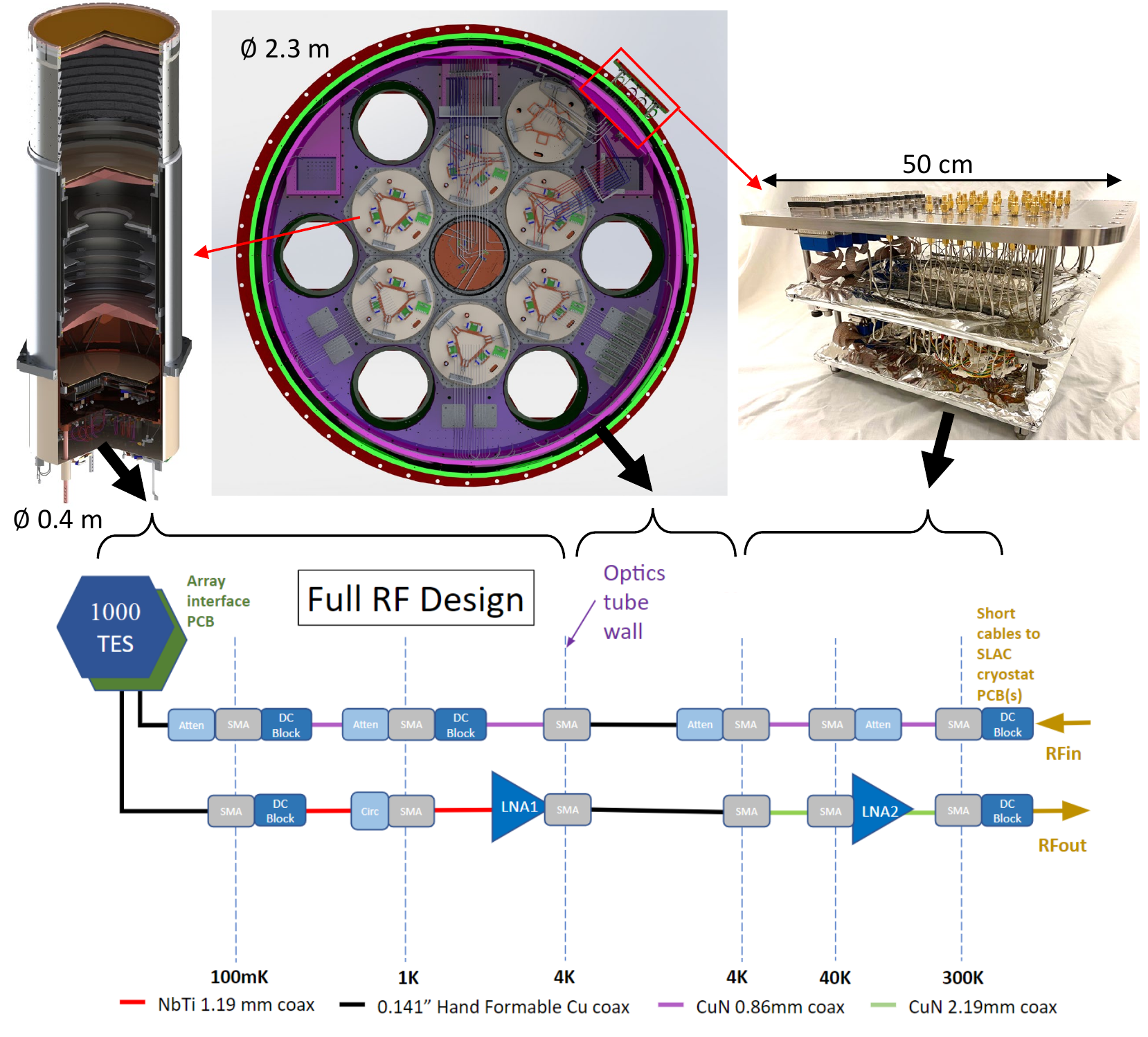}
    \caption{The $\mu$Mux readout technology\cite{sathyanarayana/etal:2020} implemented in the LATR. Pictures in the top row shows the three main parts where the $\mu$Mux system is implemented. From left to right, they are OTs, the cryostat 4\,K cavity, and the URH. More details of each of the three parts are introduced in subsequent sections. Red arrows point to where OTs and the URH are in the cryostat. Black arrows show where each part belongs in the $\mu$Mux readout architecture. The lower diagram shows the overall structure of the $\mu$Mux readout system. RF signals that drive the cold readout are sent in from the RF-in port, attenuated at several stages before arriving at the detector arrays at 100\,mK. The RF signals carry the TES signal out of the cryostat; they are amplified at two cold stages before emerging at 300~K. The color-coded lines show different coaxial cables types required at different stages for noise and thermal isolation considerations. The temperature stages are annotated by vertical dashed lines.}
    \label{fig:rf_chain}
\end{figure}

The cool-down time for each temperature stage changes with the configuration of the LATR. Intuitively, the more thermal mass and the more thermal loading leads to longer cool-down time. Currently, the most comprehensive test we have performed is the configuration with 3 window+filter sets, one OT, and URH.

During cool-downs, temperatures of different stages as a function of time are recorded by the observatory control system (OCS) software system\cite{koopman/etal:2020}. Figure~\ref{fig:cool-down_curve} shows the cool-down curve from one dark configuration and the 3-window (1-OT) configuration. In both configurations, the 80\,K stage cooled down to the base temperature within 3 days, and the 4\,K stage cooled within 4 days. After the 4\,K stage reached its base temperature, the dilution refrigerator was turned on, and cooled the 1\,K and 100\,mK stages to base temperatures within hours (Figure~\ref{fig:cool-down_curve}). The 40\,K stage, especially the 40\,K filter plate, takes 5 days to cool because of its long distance to the PT420 cold heads. Figure~\ref{fig:cool-down_curve} also shows that the 1-OT configuration cools slower compared to the dark configuration because of the addition of 3 window+filter sets, one OT, and one URH. From this measurement, we anticipate the fully-equipped LATR will cool down within 3 weeks as predicted by  thermal modeling.\cite{coppi/etal:2018}

\section{DETECTOR AND READOUT VALIDATION}
\label{sec:det_rd_validation}
SO utilizes TES detector technology with two architectures.  For 90, 150, 220, and 280\,GHz frequency bands, horn-coupled ortho-mode transducers (OMT) are used based on their performance in Atacama Cosmology Telescope (ACT)\cite{Henderson2016} and the Cosmology Large Angular Scale Surveyor (CLASS)\cite{rostem/etal:2016}.  For 30 and 40\,GHz frequency bands, the choice is lenslet-coupled, dual polarization sinuous antenna design successfully used by the POLARBEAR collaboration\cite{Suzuki2016} and the South Pole Telescope\cite{pan/etal:2018}. The hexagonal TES arrays are housed, along with the 100\,mK readout and optical coupling, in a universal focal-plane module (UFM). 
The horn-antenna OTs contain a total of 1,296 pixels that couple to 5,184 optically active TES detectors; the sinuous-antenna OTs have 111 pixels that couple to 444 optically active TES detectors\cite{Li2020, healy/etal:2020}. 

The LATR will initially deploy $\sim$40,000 TES detectors. With the detectors working simultaneously, reading them out efficiently requires a major technical development for SO, and future CMB experiments. The $\mu$Mux technology~\cite{mates2011} enables one coaxial line to read out $\mathcal{O}(10^3)$ detectors. In this architecture, each detector on a common bias line is read out by a unique resonator with a frequency between 4\,GHz and 6\,GHz. The multiplexing factor is more than 10 times that of previous technologies, such as time-domain multiplexing\cite{Henderson2016} and frequency-domain multiplexing\cite{bender/etal:2020}.

\begin{figure}[b]
    \centering
    \includegraphics[width=\textwidth]{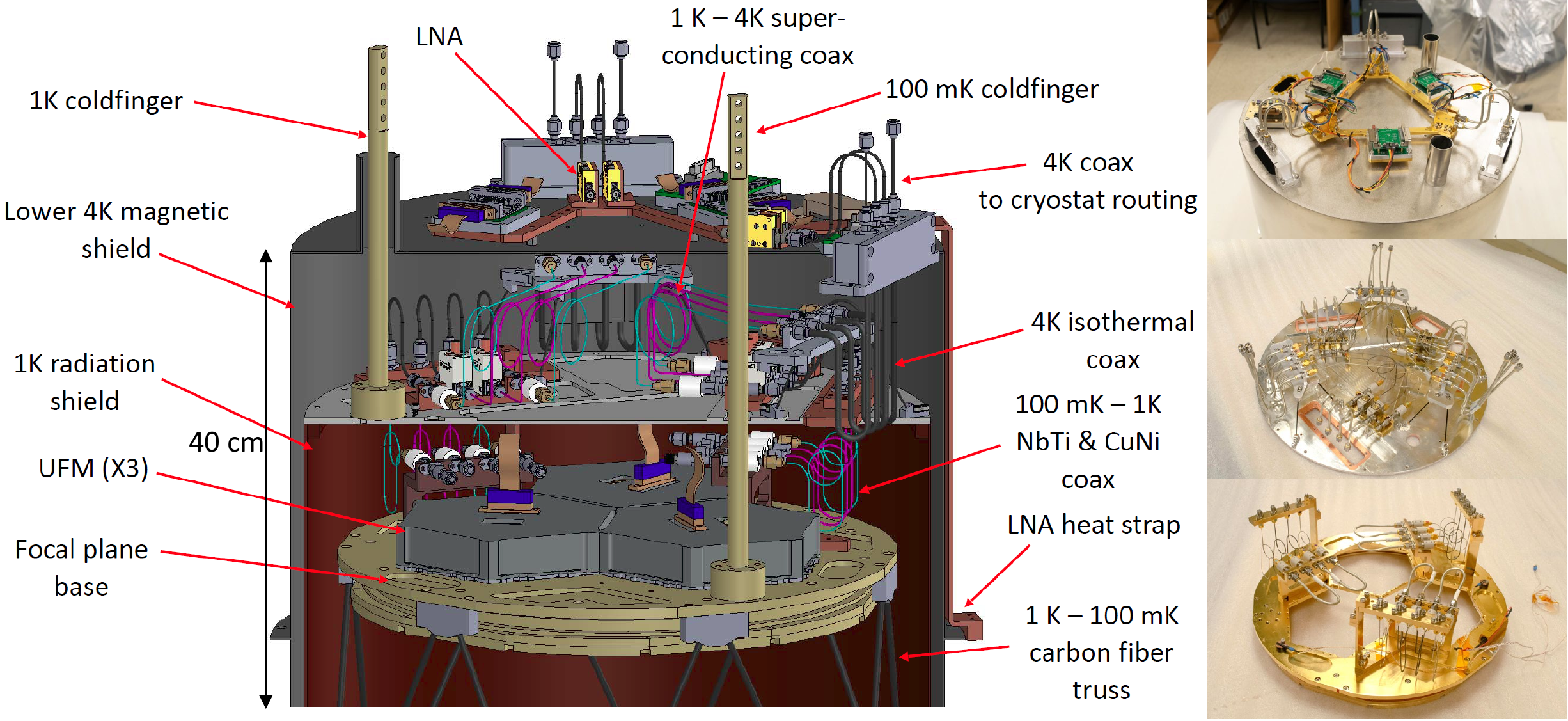}
    \caption{OT $\mu$Mux Implementation. The left part of the figure shows a detailed rendering of one OT in Figure~\ref{fig:OT_cutaway}. This rendering emphasizes the RF parts for the $\mu$Mux readout system in one OT. From bottom up, 100m\,K stage, 1\,K stage, and 4\,K stage are shown sequentially with their supporting structures and shields. Key features are also annotated. On the right part, photos of the real assembled stages are shown, following the same sequence as 100\,mK, 1\,K, and 4\,K from bottom up. Note that no UFMs were installed in the 100\,mK photo.}
    \label{fig:ot_rf}
\end{figure}

The RF architecture has been designed to achieve low noise~\cite{sathyanarayana/etal:2020}. 
Briefly, on the RF-in side, a series of attenuators lower the 300\,K thermal noise and bring the probe tones to an appropriate power at the $\mu$Mux resonators. On the RF-out side, the signals are amplified at 4\,K and 40\,K stages by low-noise amplifiers (LNAs). 
In addition, DC blocks are designed in the system for electrical and thermal considerations; circulators are also installed on the RF-out side to minimize RF reflections. Figure~\ref{fig:rf_chain} shows the one RF channel for the $\mu$Mux readout system and its implementation in the LATR.

\subsection{Microwave-multiplexing in the LATR}
\label{subsec:umux_latr}

\begin{figure}[b]
    \centering
    \includegraphics[width=\textwidth]{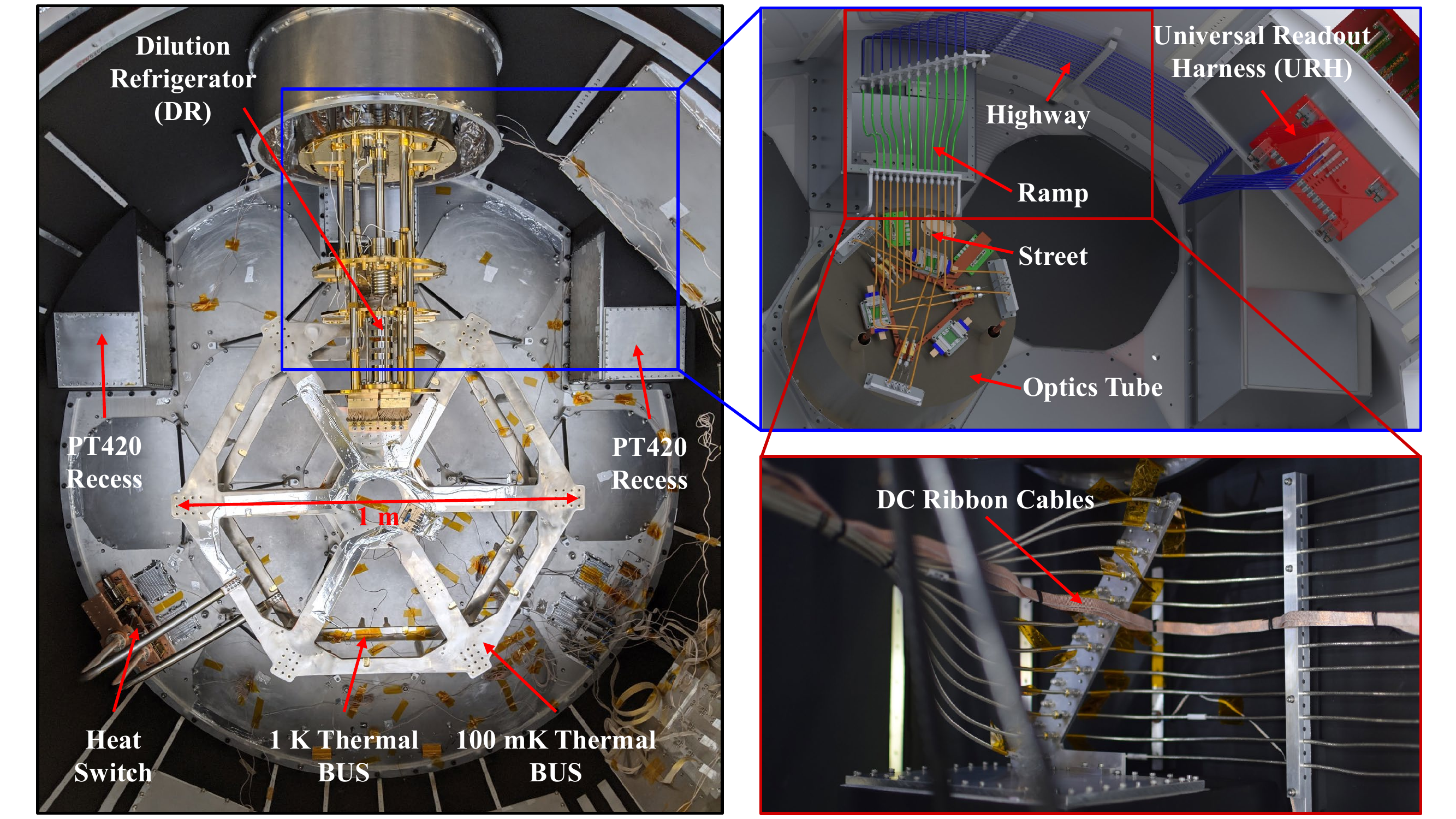}
    \caption{The $\mu$Mux cable routing in the LATR 4\,K cavity. The photo on the left shows the overview of the 4\,K stage cavity. The 1\,K and 100\,mK thermal buses, the DR, and the heat switch are shown on the 4\,K stage.  The major components are labeled along with a scale of 1\,m shown in the photo. A zoom-in plot on the top right shows the design of the coaxial route for one OT. Other isothermal 4\,K coaxial cable and the DR have been hidden for clarity. The highways are in blue, the ramps are in green, and the streets are in orange (see more descriptions in the text). Note the relatively short, simple runs of the ramps facilitate the installation and removal of optics tubes. Additionally, the connection to the URH as installed is not as angular as in this rendering. The bottom right photo shows another zoom-in photo of the 4\,K isothermal cables installed in the LATR.}
    \label{fig:4k_rf}
\end{figure}

In the LATR, each OT carries three UFMs which need up to six RF channels (twelve RF ports). In addition to the RF cables, auxiliary DC wires are also needed for detector and amplifier biasing and flux ramping\cite{sathyanarayana/etal:2020}. Due to space limitations, one challenge of the optics tube design is the routing of the readout cabling from the three UFMs at 100\,mK to the 4\,K magnetic shielding (Figure~\ref{fig:ot_rf}). To simplify the routing between isothermal components, hand-formable 3.58\,mm copper coaxial cables are
used.\footnote{Mini-Circuits, website: \href{https://www.minicircuits.com/}{https://www.minicircuits.com/}}
To connect RF readout components at different temperatures (4\,K--1\,K and 1\,K--100\,mK), semi-rigid cables are employed.  
CuproNickel (CuNi)\footnote{COAX CO., LTD., website: \href{http://www.coax.co.jp/en/}{http://www.coax.co.jp/en/}} cables with 0.86\,mm diameter are used for RF-in to control the attenuation between subsequently colder temperature stages as well as limit thermal loading. For the RF-out lines, superconducting 1.19\,mm Niobium Titanium (NbTi) cables maximize signal-to-noise, yet still have good thermal isolation properties.  Both types of semi-rigid cables are bent into loops for strain relief.  DC blocks are implemented for additional electrical and thermal isolation.  Attenuators on the input lines both at 100\,mK and 1\,K reduce noise while still providing enough power to drive the resonators. 
Each 4\,K RF output line has an LNA mounted on the back of the magnetic shield.  To reduce temperature rise due to the power generated by the LNAs ($\sim$\,5\,mW each), the LNAs are mounted on a common copper plate that has a copper strap routed down the side of each tube to the 4\,K plate. A rendering and photos of the optics tube RF design are shown in Figure~\ref{fig:ot_rf}.

Outside of the OTs, all the coaxial cables and the DC wires need to leave the cryostat at four designated ports. 
Special care should be taken in routing the coaxial cables to satisfy space constraints while minimizing interference with other OTs. 
Following this philosophy, we distribute the majority of the cable routes along the inside wall of the 4\,K shell. 
Analogous to the civil road system, our isothermal 4\,K coaxial cables are designed in three components: the highways, the ramps, and the streets (Figure~\ref{fig:4k_rf}).
The highways run along the inside of the 4\,K shell from one URH to a set of permanently installed bulkheads radially outwards from their corresponding OT. 
The highways are supported along their length, and constitute most of the length of the 4\,K isothermal route (up to $3$\,m). 
On the other end, each OT has an individual set of matching bulkheads which are attached to the back of the OT. The streets run from these bulkheads to penetrations in the 4\,K magnetic shielding, and are permanently installed in the OTs. This arrangement allows the complex routing to the back of the OTs finished outside the cryostat with a relatively simple interface between the highways and the streets. In between the street bulkheads and the highway bulkheads are the ramp sections: short ($\sim20$\,cm), simple pieces of coax, these are the only part of the isothermal 4\,K run that is not permanently installed, and hence are the only pieces that have to be added or removed when installing or removing an OT. The DC cables run along the isothermal 4\,K coaxial cables, simply being tied down to the coaxial highways, ramps, and streets. See Figure~\ref{fig:4k_rf} for an example of one OT's isothermal 4\,K cables.

\begin{figure}[b]
    \centering
    \includegraphics[width=\linewidth]{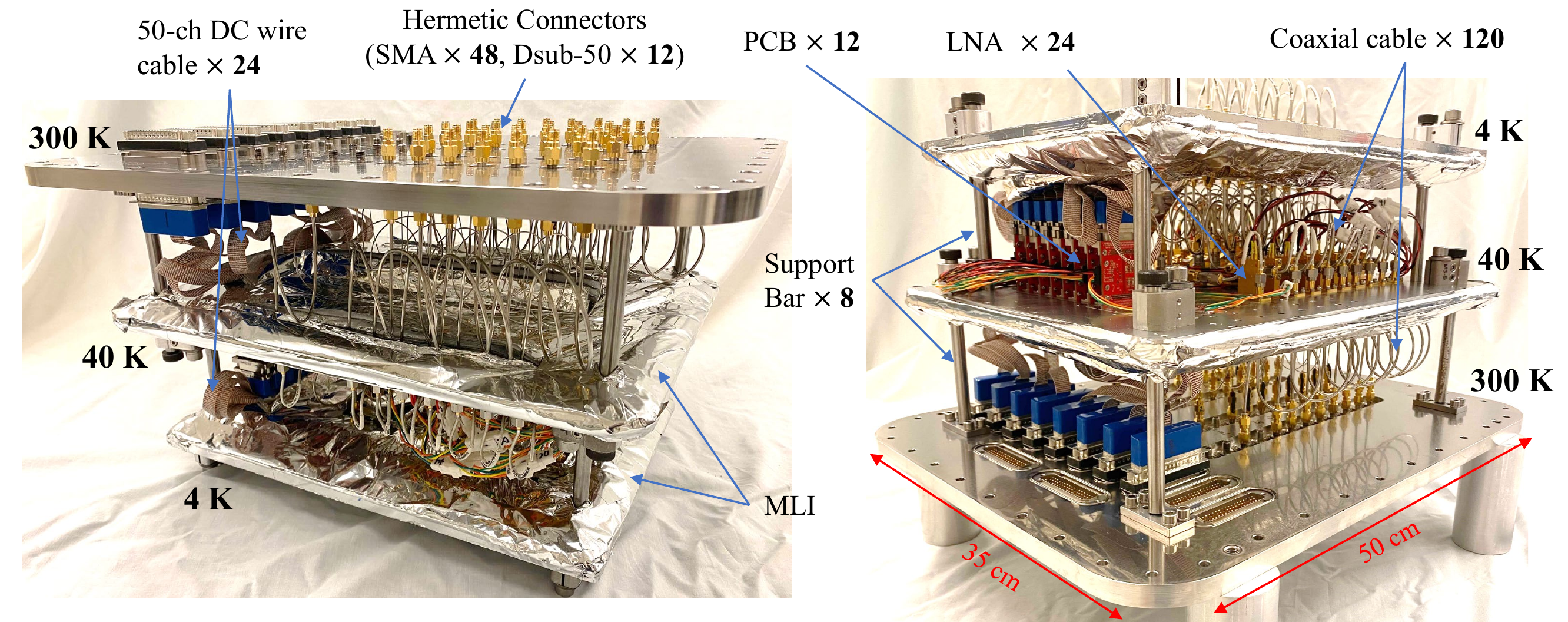}
    \caption{The SO universal readout harness (URH). Photos of one URH are presented from two perspectives. Three plates at different temperatures are annotated. Each URH can be equipped with up to 48 RF ports and 600 DC wires going from 300\,K to 4\,K. Minimal thermal loading is added from sufficient MLI installation. The 300\,K plate has 48 SMA and 12 Dsub 50-pin hermetic connectors, after which DC ribbon cables and coaxial cables continue the connections to 4\,K. RF and electronic components are also installed in-line, including low-noise amplifiers (LNAs), attenuators, and printed circuit boards (PCBs). In total, up to 24 LNAs and be installed on the RF-out side at 40\,K, accompanied by 12 PCBs to break out bias cables for the LNAs. Each of the plates are bolted onto corresponding plates in the LATR and the SATs; 8 support bars shown in the photos will be removed after installation.}
    \label{fig:urh}
\end{figure}

In the LATR, 13 optics tubes require 156 coaxial ports and 2,400 DC wires pass through different cryogenic stages with controlled thermal loading. To meet this requirement, we designed the modularized SO universal readout harness (URH) for both the LATR and the SATs\cite{ali/etal:2020}. Photos of one URH from two perspectives are shown in Figure~\ref{fig:urh}. 
One URH can contain up to 48 RF ports (120 individual coaxial cables) and 600 DC wires from 300\,K to 4\,K. In the fully-populated build, there are 96 CuproNickel semi-rigid cables (4 per RF channel). Additionally, at 40\,K there are 24 isothermal copper coax pieces. RF components including amplifiers and attenuators are also installed in situ. The LATR requires four URHs to read out all the detectors in 13 optics tubes. Given the complexity of the cable harness and the number of cables going through different stages, uncontrolled thermal loading is a major concern of the harness design. MLI sheets were tailored with minimal openings for all the cables on 40\,K and 4\,K plates. Measurements in another cryostat shows that the loading at the 40\,K stage is $\sim$\,7\,W while the loading at the 4\,K stage is $\sim$\,0.15\,W. This result ensures that the MLI design at the 40\,K and 4\,K stage successfully blocked most of the radiation loading to control the thermal loading within design. In addition, another variation of the URH was configured to readout the housekeeping data with only DC wires.

Room-temperature readout electronics---SLAC Microresonator Radio Frequency (SMuRF) Electronics\cite{henderson/etal:2018, kernasovskiy/etal:2018}---are mounted around the cryostat (see Figure~\ref{fig:latr_cut}). 
The warm electronics both drive and read back the cold readout, and ultimately detectors, as well as bias the detectors. 
The detector time streams are then packaged and stored in computing servers.

\subsection{Detector and Readout Testing Results}
\label{subsec:det_rd_results}

\begin{figure}[b]
    \centering
    \includegraphics[width=\linewidth]{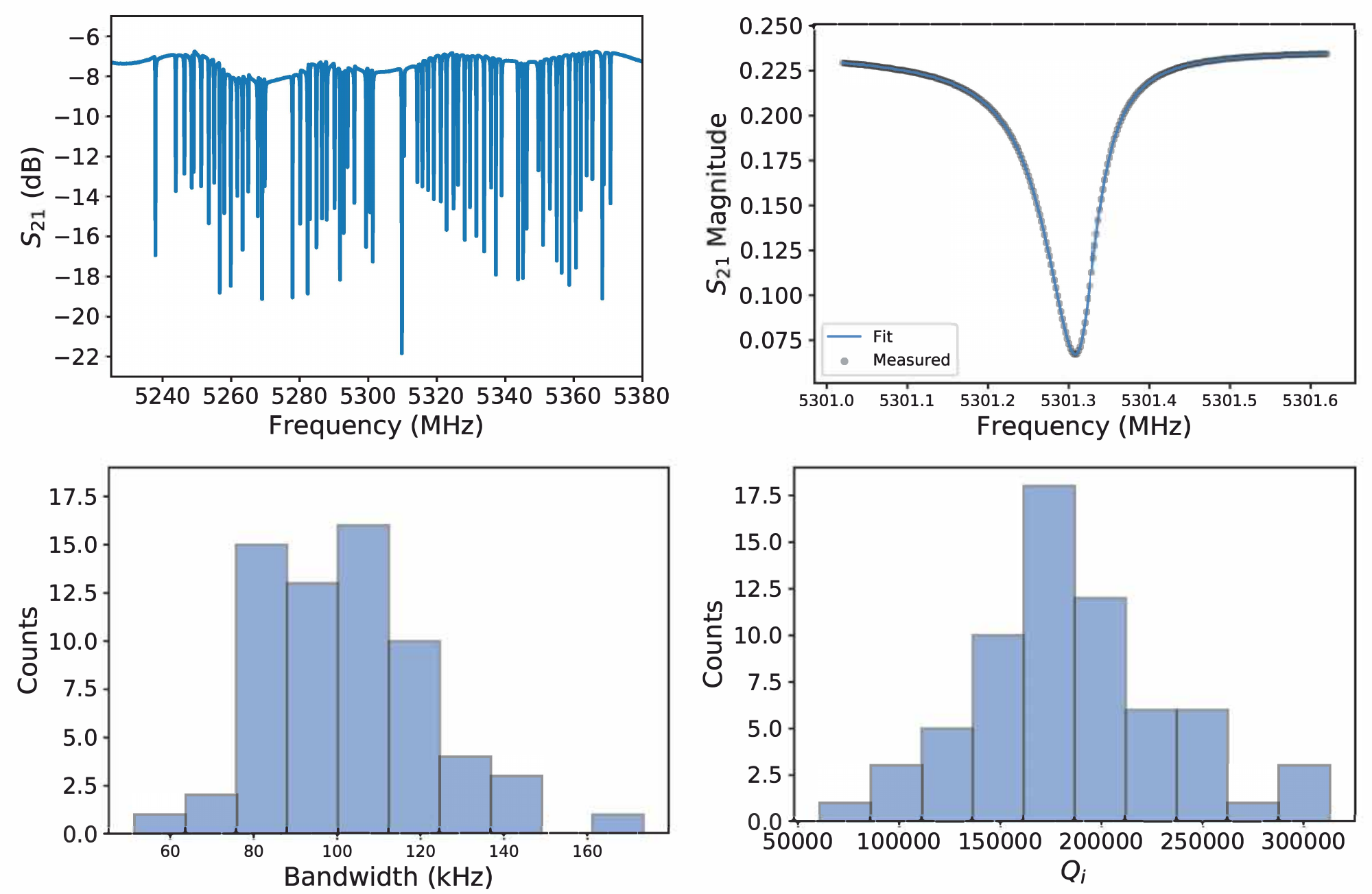}
    \caption{Multiplexer chip resonator properties as measured in the LATR.  The top left plot shows the RF transmission measurement with 65 resonance features from the multiplexer chip. The top right plot is of a single resonance at 5301\,MHz. The measured data (gray points) and the best-fit model\cite{khalil/etal:2012} (blue line) are both shown.  We fit for the internal quality-factor ($Q_i$) and calculate the bandwidth of this resonator as 2.1$\times10^5$ and 78.5\,kHz. The same analysis was conducted on other resonators in the multiplexer chip, and the histograms of the two best-fit parameters are shown in the bottom plots separately. The median resonator bandwidth is approximately 100~KHz, and the internal quality factors are generally greater than $10^5$. These results agree with measurements made in development cryostats\cite{dober/etal:2020}. }
    \label{fig:resonator_dip}
\end{figure}

Before integrating a UFM, we first tested the readout system with a demonstration assembly (SPB-3p-D-UHF), consisting of a single multiplexer chip~\cite{dober/etal:2020} and several prototype TES detectors.  The multiplexer chip contains 65 $\mu$Mux channels. In each $\mu$Mux channel, a unique RF resonator is inductively coupled to a superconducting quantum interference device (SQUID); in six of those channels, the SQUID was inductively coupled to a prototype TES detector. The assembly was cooled below the critical temperature of the detectors. 
Time streams were measured from the $\mu$Mux readout system to demonstrate the performance of it implemented in the LATR. The test was conducted in the dark configuration without the prototype TES detectors receiving optical signals. 

\begin{figure}
    \centering
    \includegraphics[width=0.7\linewidth]{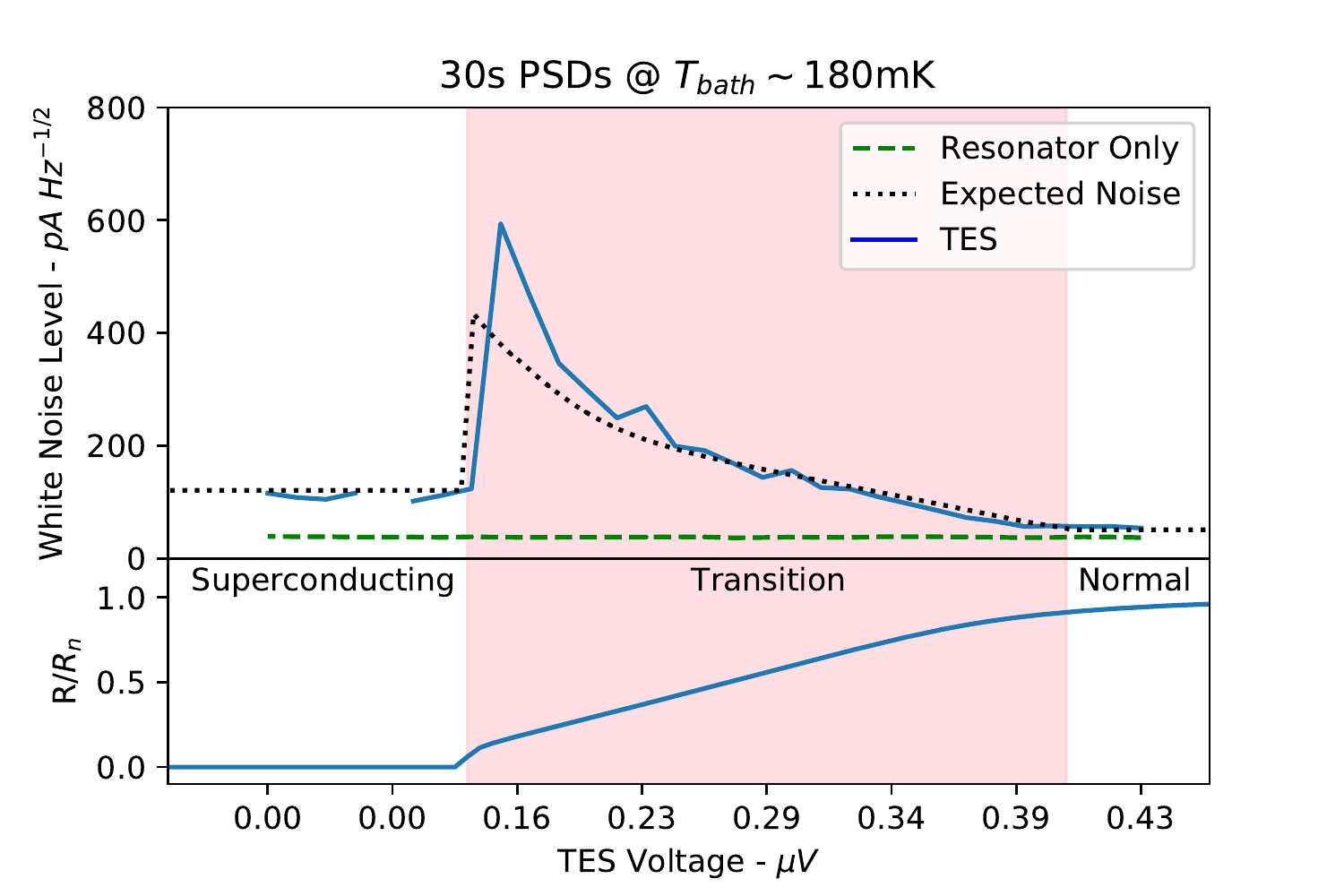}
    \caption{White noise values based on power spectral density of 30\,s time streams versus bias voltage across one prototype TES.  While detector signals are ultimately calibrated and converted to power, current through the prototype TES is what the $\mu$MUX readout system measures as the detector signal as shown in the plot. The upper panel shows the measured white noise level of one prototype TES and the median of 57 $\mu$Mux-only channels.  The comparison shows that the readout noise is subdominant to the inherent detector noise, especially in the transition state.  We also show that the expected detector noise, which is derived from measured prototype TES parameters, like $T_c\sim190$\,mK, $G\sim 11.4 \times 10^{-10}$\,W/K, $R_n\sim 8$\,m$\Omega$, and responsivities calculated from a set of IV curves. The bottom plot shows the prototype TES resistance as a fraction of normal resistance ($R_n$) as measured in the IV curves. Both panels were divided into three vertical parts illustrating the three states---superconducting, transition, and normal---of the prototype TES detector under different bias voltages. The bias voltage decreases to zero when the prototype TES enters superconducting.}
    \label{fig:noise_bias}
\end{figure}

Measurement of the 65 $\mu$Mux channel resonators are presented in Figure~\ref{fig:resonator_dip}, including RF transmission and the histograms of the resonance parameters. 
The top left panel shows the transmission across the multiplexer chip band. The top right panel shows the transmission of one resonance as well as the fit to a model\cite{khalil/etal:2012}. In this example, the resonator has an internal quality factor ($Q_i$) of $2.1\times10^5$ and a resonator bandwidth of 78.5\,kHz. 
The $Q_\mathrm{i}$ is important to the readout noise and we demonstrate here that similar properties are maintained within the LATR, compared to the properties measured in testing cryostats. 
The bottom row shows histograms of the best-fit\cite{khalil/etal:2012} bandwidth and  $Q_\mathrm{i}$ of the resonators in the multiplexer chip. The median resonator bandwidth is 103 kHz, close to the expected value of 100~kHz. The median internal quality factor $Q_\mathrm{i}$ is greater than $10^5$, as has typically been measured with these types of multiplexer chips~\cite{dicker/etal:2018}.

Time streams were then read out from prototype TES detectors and $\mu$Mux-only channels (with no detectors). White noise levels were then fit from the time streams. Figure~\ref{fig:noise_bias} shows the result of one prototype TES detector and the median of 57 $\mu$Mux-only channels.
These data were taken with $T_\textrm{bath}\sim 180$\,mK instead of the $\sim 100$\,mK fiducial operating temperature because of the higher $T_c \sim 190$\,mK of the prototype TES detectors, and to allow greater detector stability when moving through the entire transition. The prototype TES detectors in this testing setup include an extra inductor that limits stability low in the transition.
The TES white noise level is presented as a function of its bias voltage. At high $V_\textrm{bias}$, the prototype TES is normal---the high voltage resistively heats the detector above its $T_c$.  As this power decreases, the detector starts transitioning from normal to superconducting. 
In this transition the TES resistance---and so current through the prototype TES circuit---is highly sensitive to small changes in temperature.  Thus detector sensitivity, and also thermal noise pickup, is the greatest in this region. 
Figure~\ref{fig:noise_bias} shows that the readout noise is subdominant to the inherent detector noise during the transition state.
The other five prototype TES detectors in the demonstration assembly show similar results. The agreement between measured noise and expectations demonstrates that implementing the $\mu$Mux technology in the LATR---with the complexity of the RF chain design---does not degrade the performance compared to the results measured from development cryostats.

\section{Conclusion}
\label{sec:conslusion}

The large aperture telescope receiver (LATR) in the Simons Observatory (SO) fully utilizes the 1.7\,m telescope focal plane. The LATR contains five cryogenic stages, including 80\,K, 40\,K, 4\,K, 1\,K, and 100\,mK stage. Various cryogenic components, including optical filters and cryogenic lenses, are installed on the cryogenic stages to achieve designed thermal and optical performances. At the 100\,mK stage, the LATR is capable of cooling $\sim$\,60,000 transition edge sensors (TES)\cite{irwin/hilton:2005}. The 100\,mK operation temperature ensures the detector noise being sub-dominant compared to incoming photon shot noise. The detectors and the $\le$\,4\,K cold optics are packaged in modularized OTs. The modularized design facilitates installing/removing and also controls the relative positions of the optics elements to high accuracy. The fully-equipped LATR has 13 optics tubes. Acquiring data out of the cryostat requires hundreds of coaxial cables and thousands of cryogenic wires. An organized $\mu$Mux readout system is designed and being implemented in the LATR.

The LATR has been delivered and extensively tested. We started with the simplest configuration and gradually added in more factors. Currently, the LATR has been tested with 3 optical windows+filter sets, one OT installed, and two feedthroughs installed (one URH for detector readout and another one for housekeeping readout). All of the five cryogenic stages cool to their base temperatures within five days, with anticipated thermal loading and thermal gradient on different temperature stages. Extrapolating from the current results, the LATR will be able to cool all TES detectors $<$\,100\,mK as required when all the 13 OTs are installed.

The LATR uses $\mu$Mux system to readout the TES detectors. The readout system will enable one coaxial RF channel to read out $\mathcal{O}(10^3)$ TES detectors. The entire RF channel covers all temperature stages in-and-out and spans up to 5\,m in length. All coaxial cables have been designed and some have been mechanically implemented in the LATR with minimal interference for OT installation and removal. Tests on a single multiplexer chip with several prototype TES detectors have shown consistent results as obtained from test beds, showing no performance degradation from $\mu$Mux technology implemented in the LATR.

The development of the LATR offers critical information and experience on sub-100\,mK cryostats at this unprecedented size. The SO LATR design will be a valuable reference for other experiments, including CCAT-prime\cite{vavagiakis/etal:2018} and CMB-S4\cite{s4tb17, s4sb16}.

\acknowledgments 
 
This work was funded by the Simons Foundation (Award \#457687, B.K.) and the University of Pennsylvania. ZX is supported by the Gordon and Betty Moore Foundation. 

\bibliography{main} 
\bibliographystyle{spiebib} 

\end{document}